\documentclass[superscriptaddress,amsmath,twocolumn,amssymb,prd,preprintnumbers,showpacs]{revtex4}
\usepackage{amsmath,amssymb}
\usepackage{axodraw}
\usepackage{amsopn}
\usepackage{cleveref}
\usepackage{simpler-wick}
\usepackage{float}
\usepackage[cmtip,arrow]{xy}
\usepackage{pb-diagram,pb-xy}
\usepackage{slashed}
\usepackage{graphicx}

\begin{document}
\title{Optical theorem and indefinite metric in $\lambda\phi^4$ delta-theory}
\author{Ricardo Avila}
\email[Electronic mail: ]{raavilavi@gmail.com  }
\author{Carlos M. Reyes }
\email[Electronic mail: ]{ creyes@ubiobio.cl  }
\affiliation{Centro de Ciencias Exactas, Universidad del B{\'i}o-B{\'i}o, Casilla 447, 
Chill\'an, Chile.}
\begin{abstract}
A class of effective field theory called delta-theory, which improves 
ultraviolet divergences in quantum field theory, is considered. We focus on a scalar 
model with a quartic self-interaction term and construct the delta theory
by applying the so-called delta prescription. 
We quantize the theory using field variables that diagonalize the Lagrangian,
which include
a standard scalar field and a ghost or negative norm state. As well known, 
  the indefinite metric may lead to the loss of
unitary of the $S$-matrix. We study the optical theorem and check the validity of
the cutting equations 
for three processes at one-loop order, 
and found suppressed violations of 
unitarity in the delta coupling parameter
of the order of $\xi^4$.
\end{abstract}
\maketitle
\section{Introduction}
Initially, the formulation of delta theories was proposed
 to include new local gauge symmetries in non-abelian gauge theories~\cite{Delta_Gauge}. 
The Batalin-Vilkovisky quantization method for gauge 
theories provided a basis to derive delta theories 
through a classical constraint 
of the equations of motion~\cite{BV,D-Alfaro}. 
Some years later, the formalism was applied to the gravitational field to
explain the accelerated expansion of the universe~\cite{DG}
and to study some
formal aspects of quantum field theories~\cite{DQ}. 
An appealing property 
of quantum delta-theories is the possibility to suppress the radiative corrections 
 beyond one-loop order, thereby improving 
the convergence of the perturbative series.
However, the theory produces a ghost or a negative norm state
in the Hilbert space that might lead to the loss of unitarity. Despite this, the 
 model has found many applications in gravity, including
  dark energy~\cite{DG, DE}, dark matter~\cite{DM}, and 
 recently for cosmological fluctuations~\cite{CF}. 

The concept of an indefinite metric plays an essential role in quantum 
field theories~\cite{Dirac}. In gauge theories, it
reflects the redundant degrees of freedom that later become 
necessary for covariant quantization and 
prove the Ward identities. In 
electrodynamics, one can get rid of the spurious degrees 
of freedom by following the Gupta-Bleuler formalism. 
Moreover, and as is well known, higher time derivative theories
 can lead to an indefinite metric~\cite{PU}. 
 
 Lee and Wick studied indefinite metric theories and proved 
 that the $S$-matrix could be defined unitary by restricting the 
 asymptotic space~\cite{LW}. They postulated that the negative 
 metric fields decay so fast that they never appear as an \emph{in} 
 or \emph{out} states in the asymptotic Hilbert space. 
 Cutkosky proposed a covariant formulation based on 
 non-standard analyticity properties of 
 amplitudes with extended cutting rules~\cite{Cut}. 
 Lee-Wick theories have attracted a lot of attention in higher 
 derivative extensions to the standard model since they allow 
 to soften ultraviolet divergences and to solve the hierarchy 
 problem~\cite{G}. Several quantum field theory models preserve 
 unitarity under the application of the Lee-Wick formulation~\cite{Ap-LW}, 
 even in the presence of Lorentz violation~\cite{tree,unitarity}. Recently
  it has been proposed a modern formulation of Lee-Wick theories based on 
  Wick rotated Euclidean theories~\cite{P-A}.

In this work, we focus on a delta theory constructed from a $\lambda \phi^4$ 
self-interacting scalar model. We quantize the theory 
using field variables that diagonalize the Lagrangian, identify the ghost and define the physical
asymptotic Hilbert space. As a central part of the work, we study the unitarity of 
the $S$-matrix by employing the techniques of cutting diagrams within the optical theorem.

The organization of the paper is as follows. In Sec.~\ref{SecII}, we 
construct the scalar delta model. We quantize the theory and test the 
property that radiative corrections live at one-loop order. In Sec.~\ref{SecIII}, we 
diagonalize the Lagrangian and find the propagators for the standard particle and 
negative norm state. We also prove that the Hamiltonian is stable. 
Finally, in Sec~\ref{SecIV}, we explore the one-loop unitarity of the model. 
We show that violations of unitarity are present, however suppressed by 
the delta coupling parameter to
 the order of
$\xi^4$. 
\section{ Basics} \label{SecII}
In this section, we construct the delta theory and
analyze its quantum corrections. 
\subsection{The scalar $\lambda\phi^4$ delta-theory} 
Consider the scalar Lagrangian 
\begin{align}\label{L0}
\mathcal{L}_0=\frac{1}{2}\partial_\mu \phi  \, \partial^\mu \phi-\frac{1}{2}m^2
\phi^2-\frac{\lambda\phi^4}{4!}\;,
\end{align}
with mass $m$ and coupling constant $\lambda$.  

Now, we follow the delta prescription 
which consist to add to this Lagrangian the effective term
\begin{align} \label{L1}
\mathcal{L}^{\prime}(\phi, \eta)=\mathcal{L}_0+\xi \left[\frac{ 
 {\delta}\mathcal{L}_0 }
{\delta \phi (x)} \right] \eta(x)\;,
\end{align} 
where the field $\eta(x)$ is a new degree of freedom 
which has been coined the delta field~\cite{Delta_Gauge},
and $\xi$ is a small parameter that may 
be seen to arise from a more fundamental theory.

The delta-Lagrangian is
\begin{eqnarray}\label{lag_phi-eta}
\mathcal{L}' (\phi, \eta)&=&\frac{1}{2}\partial_\mu\phi \partial^\mu\phi-\frac{1}{2}
m^2\phi^2     -\frac{\lambda\phi^4}{4!} +\xi \partial_\mu\phi\partial^\mu \eta  \nonumber  
  \\ &&-\xi m^2\phi  \eta
-\xi \frac{  \lambda}{3!}\phi^3{\eta}\;.
\end{eqnarray}
where we have employed
\begin{align}\label{variation}
\frac{  {\delta}\mathcal{L}_0 }{\delta \phi (x)}=  -\partial^\mu\partial_\mu\phi  
-m^2\phi  -\frac{\lambda\phi^3}{3!} \;.
\end{align}  
The free equations of motion for the fields are
\begin{eqnarray}\label{free-eq}
\Box\phi+m^2\phi&=&0\;,
\nonumber \\
\Box\eta+m^2\eta&=&0\;.
\end{eqnarray}
The solutions correspond to the standard relation 
$p^2=m^2$ for both fields. This is a notable difference with respect to 
the gravitational sector. For gravity, due to nonlinearities and higher derivatives 
terms the delta-metric field satisfies a different equation of 
motion providing new solutions~\cite{DG}.
\subsection{Fields and propagators} 
The fields can be expanded as follows
\begin{align}\label{fields}
&\phi( \vec x,t)=\int \frac{d^3 \vec p}{(2\pi)^3}   \frac{1}{ \sqrt{2E_{ p} } }
 \left(a_{\vec p}  \,e^{ -  {\rm i} p\cdot x } + a^{\dag}_{\vec p} \,e^{ {\rm i} p\cdot x}\right)_{p_0=E_{ p}} \;,
 \\ \label{fields2}
&\eta(\vec{x},t)=\int \frac{d^3\vec p}{(2\pi)^3}\frac{1}{\sqrt{2E_{ p}}}
\left(c_{\vec p}\,e^{- {\rm i} p\cdot x}+c^{\dag}_{\vec p}\,e^{{\rm i} p\cdot x}\right)_{p_0=E_{ p}}\;,
\end{align}
with $E_{ p}=\sqrt{\vec{p}^2+m^2}$.

To find the Hamiltonian
we follow the canonical formulation. The 
canonical conjugate momenta associated to $\phi$ and $\eta$ are given by
\begin{eqnarray}
\pi_{\phi}  (x)&\equiv&\frac{\partial\mathcal{L}}{\partial\dot{\phi}}=\dot{\phi}+\xi \dot{\eta}\;,  
\nonumber \\
{\pi}_{\eta}(x)&\equiv&\frac{\partial\mathcal{L}}{\partial\dot{\eta }}=\xi \dot{\phi}\;.
\end{eqnarray}
The Legendre transformation leads to the Hamiltonian
\begin{eqnarray}
H'&=&\int d^3x \left(  \frac{1  }  {\xi}\pi_{\phi} \pi_{\eta}-\frac{1}{2\xi^2} {\pi_{\eta}}^2+\frac{1}{2}(\vec \nabla\phi)^2
+\xi \vec \nabla\phi\cdot  \vec \nabla\eta  \right. \nonumber \\ &&\left. +\frac{1}{2}m^2\phi^2
+\xi m^2\phi \eta+\frac{\lambda}{4!} \phi^4 +\frac{\xi \lambda}{3!} \phi^3 \eta   \right)\,.
\end{eqnarray}
Now, we impose the equal time commutations relations on the field operators
as follows
\begin{eqnarray}
\left[ \phi(\vec{x},t),\pi_{\phi}(\vec{x}',t)\right]&=& {\rm i}\delta(\vec{x}-\vec{x}')\,,
\nonumber \\
\left[\eta(\vec{x},t),{\pi_{\eta}}(\vec{x}',t) \right]&=&{\rm i}\delta(\vec{x}-\vec{x}')\,.
\end{eqnarray}
Substituting the fields~\eqref{fields} and~\eqref{fields2} in the above relations
we find the nontrivial elements of the algebra of commutators
\begin{eqnarray}\label{a-comm}
[a_{\vec{p}},c^{\dag}_{\vec {p}'}]&=&(2\pi)^3\delta(\vec{p}-{\vec{p}'}) \,, 
\\ \label{a-comm2}
[ c_{\vec{p}},a^{\dag}_{\vec{p}'}]&=&(2\pi)^3\delta(\vec{p}-{\vec{p}'})   \,,
\\  \label{a-comm3}
[c_{\vec{p}},c^{\dag}_{\vec{p}'}]&=&-(2\pi)^3\delta(\vec{p}-\vec{p}')\,.
\end{eqnarray}
The minus sign in~\eqref{a-comm3} is the first indication
of a negative-norm state, which eventually leads to an indefinite metric in Hilbert space, 
as we show in the next section.

In terms of creation and annihilation
 operators the Hamiltonian is
\begin{align}
H'=\int \frac{d^3 \vec p}{(2\pi)^3}E_p\left(a^{\dag}_{\vec{p}}
a_{\vec{p}}+c^{\dag}_{\vec{p}}a_{\vec{p}}
+a^{\dag}_{\vec{p}}c_{\vec{p}}\right)\,.
\end{align}
We define the vacuum state $|0\rangle$ to be annihilated by the operators
\begin{eqnarray}\label{vacuum}
a_{\vec{p}} |0 \rangle =c_{\vec{p}} |0 \rangle=0\,.
\end{eqnarray}

Let us write the Lagrangian~\eqref{lag_phi-eta} with $\lambda=0$ as
\begin{align}
\mathcal L_{\rm{free}}= \frac1 2 \Psi^T S \, \Psi   \,,
\end{align}
defining the column field $\Psi$
\begin{eqnarray}\label{Field}
\Psi= \left( \begin{array}{c}
\phi  \\  \eta  \end{array} \right)\,,
\end{eqnarray}
and the non-diagonal matrix
\begin{eqnarray}\label{operator}
S= \left( \begin{array}{cc}
Q &\xi Q  \\ 
\xi Q &0   \end{array} \right)\,,
\end{eqnarray}
with $Q=-\Box-m^2$.

The propagator follows by considering the inverse of 
$S$. In momentum space, this is
\begin{eqnarray}\label{propagator}
P= \left( \begin{array}{cc}
0& \Delta \\ \Delta  & -\Delta  \end{array} \right)\,,
\end{eqnarray}
with
\begin{eqnarray}\label{matrix-propagator}
\Delta  &=&\frac{{\rm i}}{\xi^{2}} \int \frac{d^4p}{(2\pi)^4}
\frac{e^{-{\rm i}p\cdot(x-y)}}{p^2-m^2+{\rm i}\epsilon}    \,,
\end{eqnarray}
and where we have included the ${\rm i}\epsilon$ prescription.

By considering the definition of propagator 
\begin{eqnarray}\label{propagator1}
\Delta_{ij}(x-y)&=&\frac{1}{ \xi^2 } \langle 0|T \Psi _i (x)\Psi_j(y)  |0 \rangle \,,
\end{eqnarray}
with $ \Psi _1=\phi $ and $ \Psi _2=\eta $, from Eqs.~\eqref{a-comm}-\eqref{a-comm3}
and~\eqref{vacuum} we arrive at the same result~\eqref{matrix-propagator}.
Hence, from~\eqref{propagator1} we can write
\begin{eqnarray}
\label{P_11} \Delta_{11}(z)&=&0 \\
\Delta_{12}(z)&=&\frac{{\rm i}}{\xi^{2}}  \int \frac{d^4p}{(2\pi)^4}
\frac{e^{-{\rm i}p\cdot z}}{p^2-m^2+{\rm i}\epsilon}  \,,
\\ \label{propagator2}
\Delta_{22}(z)&=&-\frac{{\rm i}}{\xi^{2}}  \int \frac{d^4p}{(2\pi)^4}
\frac{e^{-{\rm i}p\cdot z}}{p^2-m^2+{\rm i}\epsilon}   \,,
\end{eqnarray}
with $z=x-y$. 

The corrections to each matrix elements can be find with
 the perturbative series.
We begin with the lowest order correction to $\Delta_{12}(z)$ 
\begin{eqnarray}
\delta ^{(1)}\Delta_{12}(z)&=&\langle0|T\{\phi(x)\eta(y) (- {\rm i} \lambda) 
 \int d^4w  \left(\frac{\phi^4(w)}{4!}
\right.
\nonumber \\
&+&\left. \frac{\xi \phi^3(w)\eta(w)}{3!}\right)    \}|0\rangle\;.
\end{eqnarray}
According to Wick theorem and the definition of vacuum~\eqref{vacuum}
a contraction $ \wick{ \c \phi \c \phi }$
giving a possible contribution vanishes, due to 
~\eqref{P_11}. Hence, one has that $\delta^{(1)} \Delta_{12}(z)=0$, 
and the same applies for higher 
order corrections to this element. 

\begin{figure}[H]
\centering
\includegraphics[width=0.47\textwidth]{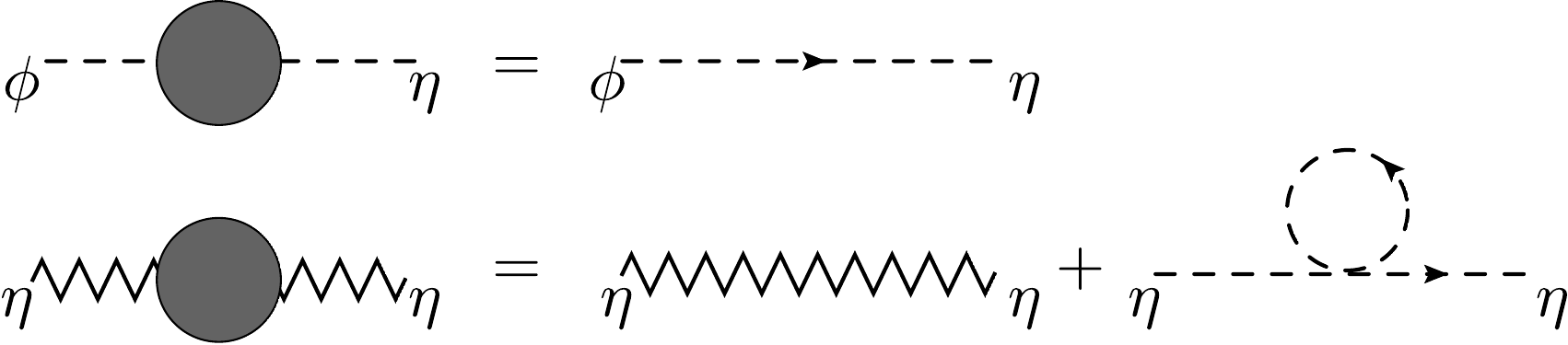}
\caption{\label{Fig1}  The 1IP diagrams for each matrix element between fields $\phi \eta$ and $\eta \eta$
of the two-point function. The propagator $ \Delta_{12}(z)= \Delta_{21}(z)$ is represented
by a segmented line and the propagator $ \Delta_{22}(z)$ by a broken line.}
\end{figure}

For the element $\Delta_{22}(z)$, the first order correction $\delta^{(1)} \Delta_{22}(z)$, 
however, is different from zero. It can be noted that 
the contraction of the external legs $\eta(x) \eta(y)$
with $\phi^3(w)\eta(w)$ gives a nonzero contribution, see Fig.~\ref{Fig1}.

However, 
at the order $\lambda^2$ one has
\begin{eqnarray}
\delta^{(2)} \Delta_{22}(z)&=&\langle0|T\{\eta(x)\eta(y) 
 \frac{(-{\rm i}\lambda)^2}{2!}\int d^4w 
d^4q 
\nonumber \\
&\times& \left(\xi^2\frac{\phi^3(w)\eta(w)}{3!} \frac{\phi^3(q)\eta(q)}{3!}
+\frac{\phi^4(w)\phi^4(q)}{(4!)^2}\right.
\nonumber \\
&+&\left.\xi\frac{\phi^3(w)\eta(w)}{3! }  \frac{\phi^4(q)}{4!} 
\right.
\nonumber \\
&+&\left.  \xi \frac{\phi^4(w)}{4!}  \frac{\phi^3(q)\eta(q)}{3!}  \right)    \}|0\rangle\,,
\end{eqnarray}
which vanishes due to the same previous arguments. 

The only non-vanishing quantum corrections for
 the matrix propagator comes at one-loop order and so, 
 we have tested one of the basic properties of delta theories
in the scalar sector. 
\subsection{Effective action} 
Consider the vacuum-vacuum amplitude for the 
Lagrangian~\eqref{lag_phi-eta} in the presence of the
currents $J$ and $\widetilde J$
\begin{equation}\label{generating_Z}
Z[J, \widetilde J]=\int \mathcal D\phi \, \mathcal D\eta \, e^{{\rm i} \int d^4x \left( \mathcal
 L'(\phi,\eta) +J \phi+\tilde J \eta \right) }\;.
\end{equation}
The equations of motion for $\phi$ are
\begin{eqnarray}\label{Eq_current}
\xi \Box\phi+\xi m^2\phi +\xi \frac{\lambda}{3!}\phi^3=\widetilde  J \;.
\end{eqnarray}
and for $\eta$
\begin{eqnarray}\label{Eq_current2}
\Box\phi+ m^2\phi +\frac{\lambda}{3!}\phi^3+\xi \Box\eta+\xi m^2\eta +\xi \frac{\lambda}{2!}\phi^2\eta=  J \;.
\end{eqnarray}
We integrate~\eqref{generating_Z} with respect to the field $\eta$, and obtain
\begin{eqnarray}\label{gen-func2}
Z[J, \widetilde J]&=&\int \mathcal D\phi  \,e^{{\rm i}\int d^4x \left(\frac{1}{2}\partial_\mu\phi\partial^\mu\phi
-\frac{1}{2}m^2\phi^2+\frac{\lambda\phi^4}{4!}+J\phi\right)}\nonumber \\
 &\times&\delta  \left(-\xi \Box\phi-\xi m^2\phi-\xi \frac{\lambda}{3!}\phi^3+\widetilde{J}  \right)\;.
\end{eqnarray}
In terms of the classical solutions $\phi_0$ of 
the equation of motion~\eqref{Eq_current}, we can expand the delta as 
\begin{eqnarray}
&&\delta \left(-\xi \Box\phi-\xi m^2\phi-\xi \frac{\lambda}{3!}\phi^3+\widetilde{J} \right)\\\nonumber
&=&\text {det}^{-1}\left(-\xi  \Box-\xi  m^2-\xi  \frac{\lambda}{2!}\phi^2\right)|_{\phi=\phi_0
}\delta(\phi-\phi_0)\;.
\end{eqnarray}
We
substitute in~\eqref{gen-func2} and integrate, which yields
\begin{eqnarray}\label{Z-gen}
Z[J, \widetilde J]&=&e^{{\rm i}\int d^4x \left(\frac{1}{2}\partial_\mu\phi_0\partial^\mu\phi_0
-\frac{1}{2}m^2\phi^2_0+\frac{\lambda\phi^4_0}{4!}+J\phi_0\right)}  \nonumber \\
&& \text{det}^{-1}\left(-\xi  \Box-\xi  m^2-\xi  \frac{\lambda}{2!}\phi^2_0 \right)\;.
\end{eqnarray}
Consider the generating function of connected Green's functions 
\begin{eqnarray}
W[J, \widetilde J]&=&-{\rm i} \, \text{ln} \, Z[J, \widetilde J] \;,
\end{eqnarray}
which in~\eqref{Z-gen} produces
\begin{eqnarray}
W[J, \widetilde J]&=&\int d^4x \left(\frac{1}{2}\partial_\mu\phi_0\partial^\mu\phi_0
-\frac{1}{2}m^2\phi^2_0+\frac{\lambda\phi^4_0}{4!}+J\phi_0\right) \nonumber \\
&+&{\rm i}\text{Tr}\, \text{log} \left(-\xi  \Box-\xi  m^2-\xi  \frac{\lambda}{2!}\phi^2_0 \right)\;.
\end{eqnarray}
We define as usual the effective action by 
\begin{eqnarray}\label{Gamma}
\Gamma[\Phi,\widetilde{\Phi}]&=&W[J,\widetilde{J}]-\int d^4x
 \left(J\Phi+\widetilde{J}\, \widetilde{\Phi} \right)  \;,
\end{eqnarray}
with the classical fields
\begin{eqnarray}
\Phi&=&\frac{\delta W}{\delta J}=\phi_0  \;, \\
\widetilde{\Phi}&=&\frac{\delta W}{\delta \widetilde{J}}=0\;.
\end{eqnarray}
Finally, we have
\begin{eqnarray}\label{Gamma_complete}
\Gamma[\Phi,\widetilde{\Phi}]&=&\int d^4x \left(\frac{1}{2}
\partial_\mu\Phi\partial^\mu\Phi-\frac{1}{2}m^2\Phi^2
+\frac{\lambda\Phi^4}{4!}\right)   \nonumber  \\
&+&{\rm i} \text{Tr}\text{ log}(-\Box-m^2-\frac{\lambda}{2!}\Phi^2) \;.
\end{eqnarray}
Comparing with the quantum correction for an arbitrary action $S$
\begin{eqnarray}
\Gamma[\Phi]_{1loop}=S[\Phi]+\frac{{\rm i}}{2}\text{Tr} 
 \text{log}\left(\frac{\delta^2S[\Phi]}{\delta\Phi\,\,\delta\Phi}\right)\;,
\end{eqnarray}
we have found for our $\lambda\phi^4$ theory
\begin{eqnarray}
\Gamma[\Phi]_{1loop}=S[\Phi]+\frac{{\rm i}}{2} \text{Tr}
\text{ log}(-\Box-m^2-\frac{\lambda}{2!}\Phi^2)\;.
\end{eqnarray}
In this way, we note that the effective action in our delta theory~\eqref{Gamma_complete} involves
just the a one-loop 
correction, see the general derivation~\cite{DQ}. Also, in comparison with the standard $\lambda \phi^4$ model
it has been amplified by a factor of two, which is
characteristic of delta theories.

An alternative demonstration can be given using 
diagrammatic arguments~\cite{Delta_Gauge}. 
Let us introduce the notation, $V_1$ for the number of vertices
associated to the $\phi^4$ interaction and $V_2$ to number of vertices 
associated to the interaction
$\phi^3 \eta$. We denote the number of internal lines 
$I_1$ associated to the propagator $\Delta_{12}= \Delta_{21}$, and 
$I_2$ associated
 the propagator $\Delta_{22}$.
 
Now, consider the general relation for a loop diagram
\begin{align}\label{topo}
L=I-V+1
\end{align}
where $L$ denotes the number of loops, $I$ the number of internal lines and $V$ the number of
vertices. Since it is not possible to have external legs $\phi$, the only vertex 
involved in the internal part of a diagram is $V_2$.  The contraction of $\eta \phi$ in $V_2$ produces a $I_1$. Hence, 
one has
\begin{eqnarray}
L&=&I_1-V_2+1\;.
\end{eqnarray}
The correspondence $I_1=V_2$ is also consequence of the contraction. 
We have that the maximum allowed number of loops is $L=1$.
\section{Physical fields and their propagators}\label{SecIII}
The diagonalization of the Lagrangian~\eqref{lag_phi-eta}
can be achieved by introducing the new fields
\begin{eqnarray}
\phi_1&=&\phi+\eta \;, \nonumber  \\
\phi_2&=&\eta\;.
\end{eqnarray}
Substituting in~\eqref{lag_phi-eta} yields the Lagrangian
\begin{eqnarray}\label{Lag2}
\mathcal{ L}&=&\frac{1}{2}\partial_\mu\phi_1\partial^\mu\phi_1-\frac{1}{2}
\partial_\mu\phi_2\partial^\mu\phi_2-\frac{m^2}{2}\phi_1^2+\frac{m^2}{2}
\phi_2^2\nonumber \\ 
 &-&\frac{\lambda}{4!}\left(  \phi_1^4-3 \phi_2^4-6\phi_1^2\phi_2^2+8\varepsilon \phi_1
 \phi_2^3\right)\;,
\end{eqnarray}
where we have absorbed $\xi$ into the field $\phi_2$ and $\varepsilon=\pm 1$
depends on the sign of $\xi$.

We write the new fields as
\begin{align}\label{fields_phi1}
&\phi_1( \vec x,t)=\int \frac{d^3 \vec p}{(2\pi)^3}   \frac{1}{ \sqrt{2E_{ p} } }
 \left(b_{1\vec p}  \,e^{ -{\rm i} p\cdot x } + b^{\dag}_{1\vec p} \,e^{{\rm i} p\cdot x}\right)_{p_0=E_{ p}} \;,
 \\ \label{fields2_phi}
&\phi_2(\vec{x},t)=\int \frac{d^3\vec p}{(2\pi)^3}\frac{1}{\sqrt{2E_{ p}}}
\left(b_{2\vec p}\,e^{-{\rm i} p\cdot x}+b^{\dag}_{2\vec p}\,e^{{\rm i} p\cdot x}\right)_{p_0=E_{ p}}\;,
\end{align}
and define the new creation and annihilation
\begin{align}
b_{1\vec{p}}=a_{\vec{p}}+c_{\vec{p}} \;,
   \qquad    b_{1\vec{p}}^{\dag}=a_{\vec{p}}^{\dag}
   +c_{\vec{p}} ^{\dag}     \;,  
\\
b_{2\vec{p}}=c_{\vec{p}} \;,  
\qquad b_{2\vec{p}}^{\dag}=c_{\vec{p}}  ^{\dag}    \;.
\end{align}
in terms of the ones in~\eqref{fields} and~\eqref{fields2}.

It can be checked by employing~\eqref{a-comm}, that 
they satisfy the commutations relations
\begin{align}\label{comm1}
&\left[b_{1\vec{p}},  b^{\dag}_{1  \vec q  }    \right]=(2\pi)^3\delta(\vec{p}-\vec{q}) \;,
\\ \label{comm2}
&\left[b_{2\vec{p}},b^{\dag}_{2\vec q} \right]=-(2\pi)^3\delta(\vec{p}-\vec{q})\;,
\end{align}
with all others commutators being exactly zero.

The free Hamiltonian is found to be
\begin{eqnarray}
H=\int \frac{d^3p}{(2\pi)^3} E_p \left(b^{\dag}_{1\vec{p}}
b_{1\vec{p}}-b^{\dag}_{2\vec{p}}b_{2\vec{p}} \right)\;.
\end{eqnarray}
which can be expected due to the negative-norm state.
Moreover, as a result of our previous definition of vacuum~\eqref{vacuum},
we have
\begin{eqnarray}\label{newvacuum}
b_{1\vec{p}}|0\rangle=0\;, \\  \nonumber
b_{2\vec{p}}|0\rangle=0\;.  
\end{eqnarray}
The number operators associated to the two types of particles are
\begin{eqnarray}
N_{1\vec{p}}&=&b^\dag_{1\vec{p}}\, b_{1\vec{p}} \;,  \\ \nonumber
N_{2\vec{p}}&=&-b^\dag_{2\vec{p}} \, b_{2\vec{p}}\;.
\end{eqnarray}
In terms of the number operators, the Hamiltonian is given by
\begin{eqnarray}
H=\int \frac{d^3p}{(2\pi)^3} E_p (N_{1\vec{p}}+N_{2\vec{p}})\;,
\end{eqnarray}
which is bounded from below and 
stable when the interactions are turned on.
From the commutators~\eqref{comm1} and~\eqref{comm2},
we identify the creation operator $b^{\dag}_{1\vec{p}}$ associated to a positive metric
 and $b^{\dag}_{2\vec{p}}$ associated to a negative metric~\cite{LW}.

\begin{figure}[H]
\centering
\includegraphics[width=0.30\textwidth]{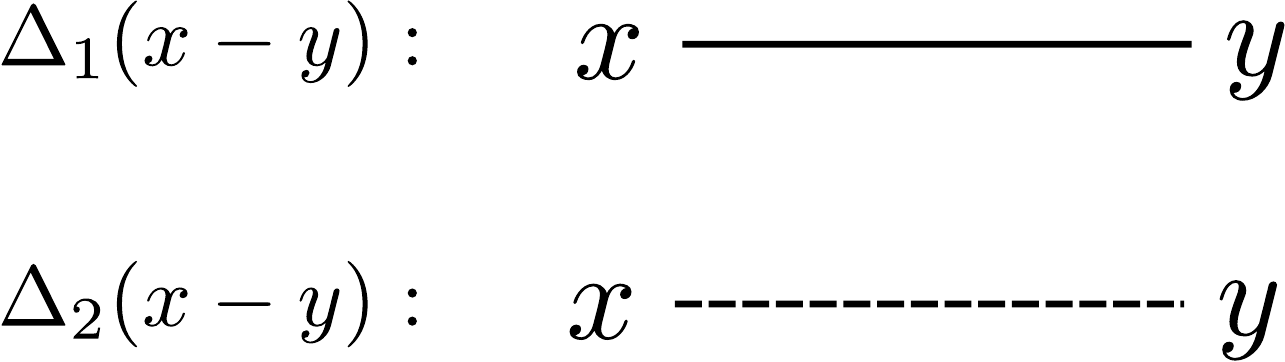}
\caption{\label{Fig2}  For the propagator $\Delta_{1}(x-y)$
we consider a normal line and for the propagator 
$\Delta_{2}(x-y)$ a segmented line.}
\end{figure}

The propagators follows by the usual definition
\begin{eqnarray}
\Delta_{1}(x-y) &=&  \langle 0|T \phi_1  (x)\phi_1(y)  |0 \rangle \nonumber \\
&=&{\rm i} \int _{\mathcal C_1}  \frac{d^4p}{(2\pi)^4}\frac{e^{-{\rm i} p\cdot(x-y)}}{p^2-m^2+{\rm i} \epsilon}\;,
\end{eqnarray}
and
\begin{eqnarray}
\Delta_{2}(x-y)&=&  \langle 0|T \phi_2  (x)\phi_2(y)  |0 \rangle \nonumber \\ 
 &=& -{\rm i} \int _{\mathcal C_1} \frac{d^4p}{(2\pi)^4}\frac{e^{-{\rm i} p\cdot(x-y)}}{p^2-m^2+{\rm i} \epsilon}\;,
\end{eqnarray}
where the contour $\mathcal C_1$ in the complex $p_0$-plane 
lies below the negative pole and above the positive one.
Both propagators as depicted in Fig~\eqref{Fig2}.

\begin{figure}[H]
\centering
\includegraphics[width=0.4\textwidth]{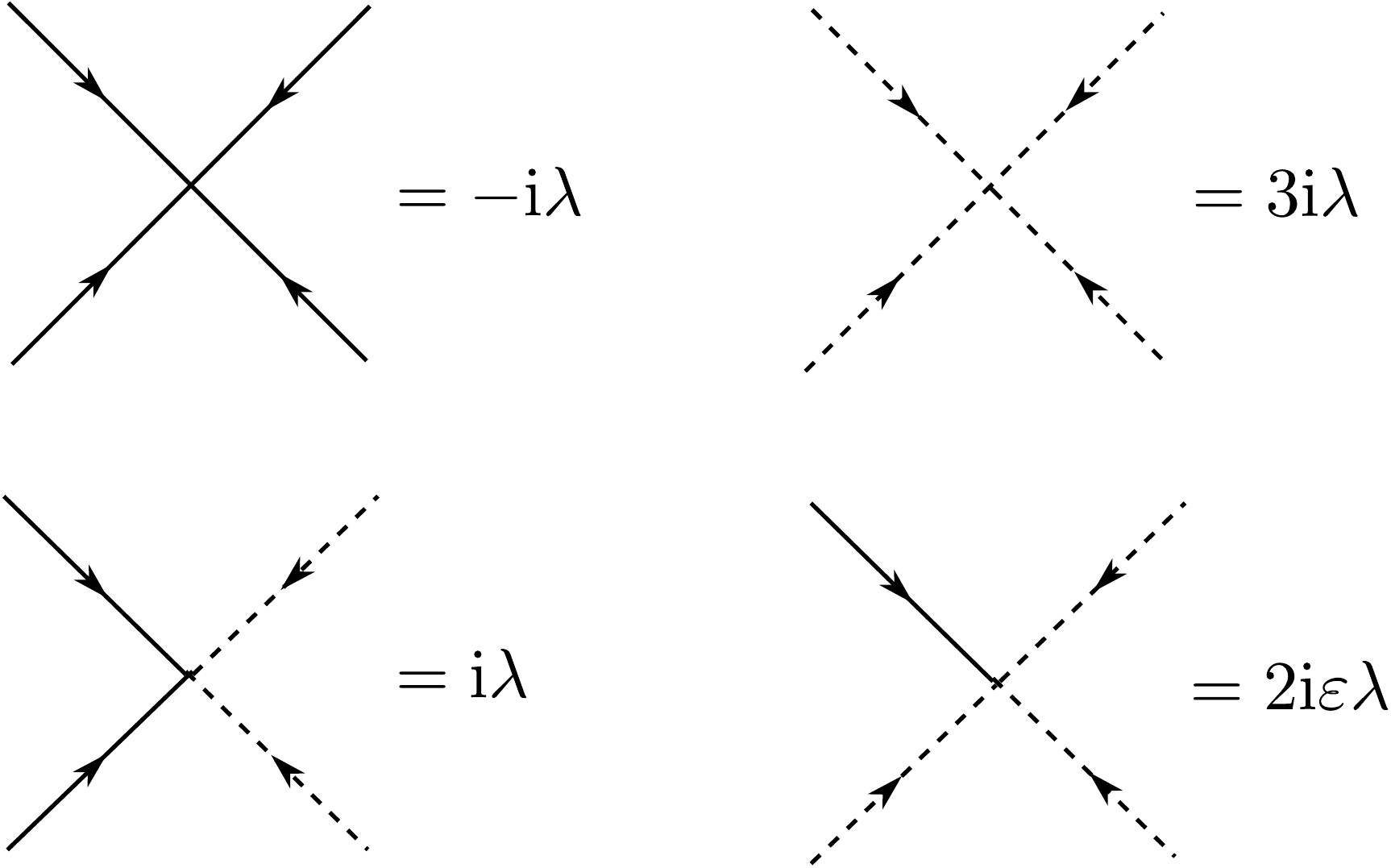}
\caption{\label{Fig3}  The four vertices in the 
diagonal base $\phi_1,\phi_2$.}
\end{figure}

We can also draw the vertices according to the interacting term 
in the Lagrangian~\eqref{Lag2}, see Fig~\eqref{Fig3}. It is convenient to mention that 
the property of having radiative correction up to one loop order is lost in this new basis. However one may expect 
that writing the original quantum corrections at a given order in terms of the new fields these are cancelled out.
\section{Perturbative Unitarity }\label{SecIV}
In this section, we study perturbative unitarity by 
verifying the optical theorem.
\begin{figure}[H]
\centering
\includegraphics[width=0.3\textwidth]{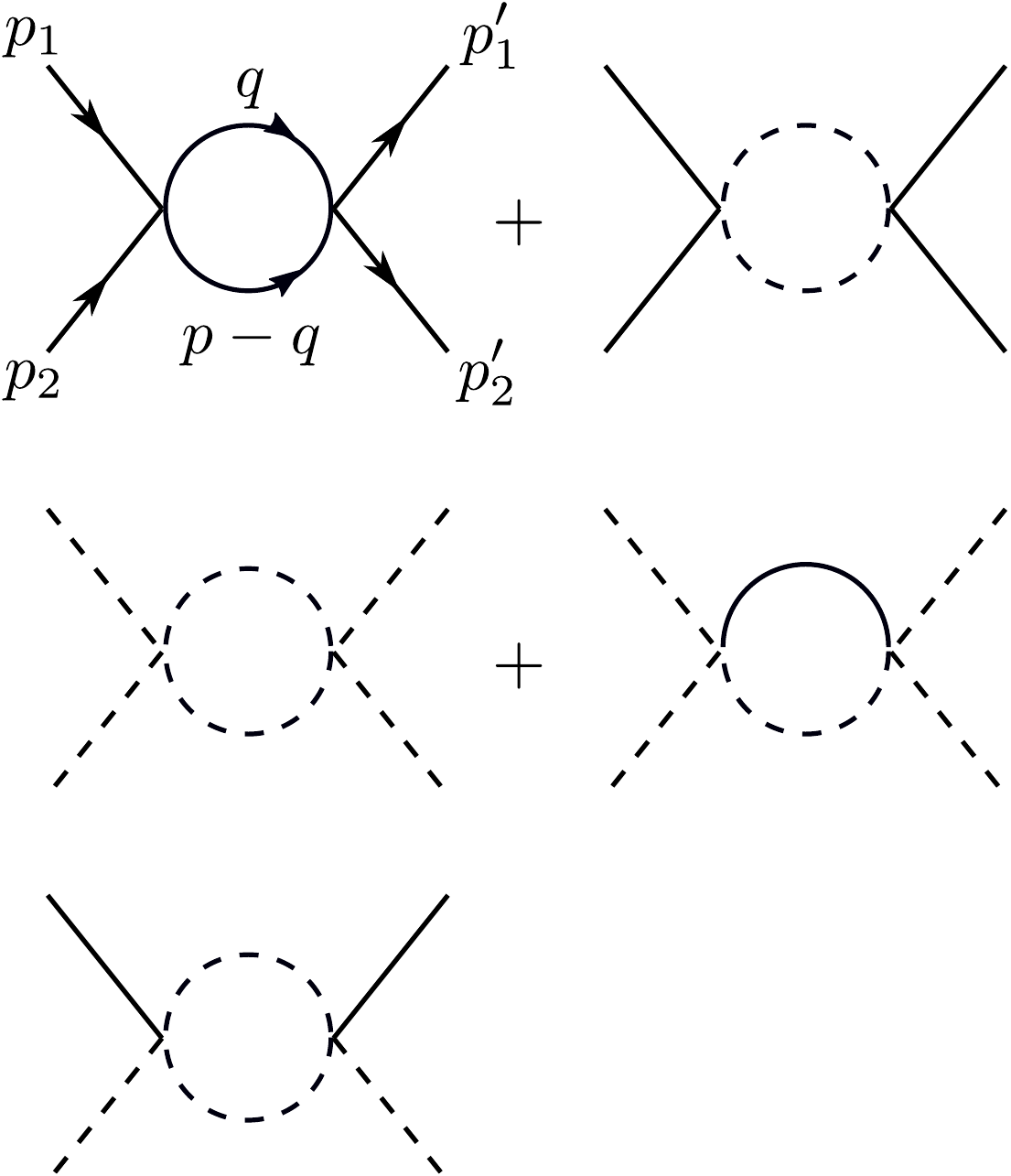}
\caption{\label{Fig4} Diagrams contributing to the 
forward scattering $p_1+p_2\to p'_1+p'_2$ at one-loop order.}
\end{figure}

In terms of the physical fields
 $\phi_1$ and $\phi_2$, we focus on the process
of two particles going to two particles 
with
initial and final momenta $(p_1, p_2)$ and $(p'_1, p'_2)$, respectively
\begin{eqnarray}
p_1+p_2\to p'_1+p'_2\;,
\end{eqnarray}
as seen in Fig.~\eqref{Fig4}
 
We can reinstate the small 
coupling constant $\xi$ for the propagator and the vertex,
so that the diagrams in the 
first line, second and third line of Fig.~\ref{Fig4} are of order
$\lambda^2 $, $\lambda^2 \xi ^4$ and $\lambda^2 \xi^2$, respectively.

The amplitudes are given by
\begin{eqnarray}\label{amplitudes}
{\rm i} \mathcal{M}_1(p)&=&\frac{1}{2}  \mathcal{M}_1^{(1, 1)}(p)+ \frac{1}{2} \mathcal{M}_1^{(2, 2)}(p) \;,
\nonumber \\
\nonumber {\rm i}  \mathcal{M}_2(p)&=& \frac{9}{2}\mathcal{M}_2^{(2, 2)}(p)+4 \mathcal{M}_2^{(1, 2)}(p) \;,
\\
{\rm i} \mathcal{M}_3(p)&=&  4 \mathcal{M}_3^{(2, 2)}(p)\;,
\end{eqnarray}
where  $p=p_1+p_2$ and we have included the symmetry factors for each diagram.
Each element is defined by
\begin{eqnarray}
\mathcal M_i^{(a, b)}(p)&=& (- {\rm i}\lambda_i)^2  \int\frac{d^4q}{(2\pi)^4}
\Delta_{a} (q)  \Delta_{b}(p-q) \,,
\end{eqnarray}
where $a,b=1,2$, $i=1,2,3$, such that $\lambda_1=\lambda $, $\lambda_2=\lambda \xi^2$, $\lambda_3=  \lambda \xi$, and 
 the propagators for $\phi_1$ and $\phi_2$
\begin{eqnarray}
\Delta_{1} (q)&=&    \frac{  \rm{i}  } {q^2-m^2+\rm{i} \epsilon}\,,
 \\
\Delta_{2} (q)&=&  \frac{ - \rm{i}}{q^2-m^2+\rm{i} \epsilon}\,.
\end{eqnarray}
We compute the imaginary part of the amplitudes by computing its discontinuity.
To begin with, let us write the loop integral
\begin{eqnarray}
\mathcal{M}_1^{(1, 1)}(p)&=& (- {\rm i}\lambda_1)^2 \int\frac{d^4q}{(2\pi)^4}\frac{\rm{i}}{q^2-m^2+{\rm i} \epsilon}\;
\\ &\times&  \frac{\rm{i}}{(p-q)^2-m^2+\rm{i} \epsilon}   \nonumber \,,
\end{eqnarray}
with two propagators $\Delta_{1} (q)$ and $\Delta_{1} (p-q)$ represented by the first diagram in 
Fig.~\ref{Fig4}.

We employ the residue theorem to compute the integral in the complex $q_0$-plane.
For this, we integrate along the contour $\mathcal C_1$ that encloses the poles of the lower half plane.
Hence, we consider
\begin{eqnarray}
&& \mathcal M_1^{(1, 1)}(p)= \lambda_1^2 
 \int\frac{d^3 \vec q}{(2\pi)^3}  \int_{\mathcal C_1}\frac{d  q_0}{2\pi} \\ &\times&\frac{1}{(q_0-E_q+ {\rm i}\epsilon )(q_0+E_q-\rm{i}\epsilon)}\;
\nonumber  \\ &\times& \frac{1}{(q_0-p_0-E_{ q-p}+{\rm i} \epsilon )(q_0-p_0+E_{q-p}-{\rm i} \epsilon)}  \nonumber \,,
\end{eqnarray}
with $E_q=\sqrt{\vec q^2+m^2}$ and the poles at
\begin{eqnarray}
q_0&=&E_q- {\rm i} \epsilon \;, \nonumber \\
q_0&=&p_0+E_{q-p}-{\rm i} \epsilon\;.
\end{eqnarray}
The integration gives
\begin{eqnarray}
&&\mathcal M_1^{(1, 1)}(p)= \lambda_1^2 \int\frac{d^3 \vec q}{(2\pi)^4}(-2\pi {\rm i})  \\ &&
\left( \frac{1  }{2E_q(-p_0+E_q-E_{q-p})(-p_0+E_q+E_{q-p}-i\epsilon
 )}   \right.  \nonumber \\ &+& \left. 
 \frac{ 1  }{2E_{q-p}(p_0+E_{q-p}-E_q)(p_0+E_{q-p}+E_q-i\epsilon
 )}   \right) \nonumber \,,
\end{eqnarray}
where we have rescaled $\epsilon$ and evaluated $\epsilon \to 0$ where it is not relevant for the discontinuity.
It is convenient to compute its discontinuity and so, we employ the relation
\begin{eqnarray}
\lim _{\epsilon \to 0^+}\frac{1}{x\pm {\rm i} \epsilon }=
\mathcal P \left( \frac{1}{x}\right)  \mp i\pi\delta(x)\;,
\end{eqnarray}
with $\mathcal P$ the principal value.
This results in the discontinuity
\begin{eqnarray}
&&{\rm Disc} \mathcal M_1^{(1, 1)}(p)={\rm i} \lambda_1^2
 \int\frac{d^3 \vec q}{(2\pi)^3}\frac{1}{2E_qE_{q-p}}
 \\ && \left({\rm i} \pi\delta(p_0-E_p-E_{q-p})    \right.\nonumber  \\  &+& \left. 
 {\rm i} \pi\delta(p_0+E_q+E_{q-p})     \right)  \nonumber \;.
\end{eqnarray}
For simplicity we consider $p_0>0$ and 
we relabel the two momenta of each propagator as $q_1$ 
and $q_2$, such that
\begin{eqnarray}
\vec {q}=\vec{q}_1\;, \\
\vec{p}-\vec{q}=\vec{q}_2 \;.
\end{eqnarray}
We can write
\begin{eqnarray}
&&{\rm Disc} \mathcal M_1^{(1, 1)}(p)={\rm i} \lambda_1^2 \int\frac{d^3\vec q_1 }
{(2\pi)^3} \frac{d^3\vec q_2 }{(2\pi)^3} 
\\ &&  \frac{ \delta^3(\vec p-\vec q_1-\vec q_2 ) }{2E_{q_1}E_{q_2}}
 ({\rm i} \pi)  \delta(p_0-E_{q_1} -E_{q_2} )  \nonumber  \;,
\end{eqnarray}
or
\begin{eqnarray}
{\rm Disc} \mathcal M_1^{(1, 1)}(p)&=&{\rm i} \lambda_1^2 \int\frac{d^4 q_1 }
{(2\pi)^4} \frac{d^4 q_2 }{(2\pi)^4}  (2\pi)^4  \nonumber\\ && \delta^4(p- q_1- q_2 )
 \delta(q_1^2-m^2)\\ &&  \delta(q_2^2-m^2) \theta(q_{01}) \theta(q_{02})  \nonumber \;.
\end{eqnarray}
Using the relation ${\rm Disc} \mathcal M=2 {\rm i} {\rm Im } \mathcal{M}$ we have
\begin{eqnarray}
2 {\rm Im} \mathcal M_1^{(1, 1)}(p)&=& \lambda_1^2 \int\frac{d^4 q_1 }
{(2\pi)^4} \frac{d^4 q_2 }{(2\pi)^4}  (2\pi)^4  \nonumber \\ && \delta^4(p- q_1- q_2 )
 \delta(q_1^2-m^2) \\ && \delta(q_2^2-m^2) \theta(q_{01}) \theta(q_{02})  \nonumber \;.
\end{eqnarray}
The second and third type of integral follow in a similar way, such to have 
\begin{eqnarray}
 \frac{{\rm Im} \mathcal M_1^{(1, 1)}(p)}{\lambda_1^2}=-\frac{ {\rm Im} 
 \mathcal M_2^{(1, 2)}(p)}{ \lambda_2^2}=\frac{ {\rm Im} \mathcal M_{1,2,3}^{(2, 2)}(p)}{\lambda_{1,2,3}^2 }\;.
\end{eqnarray}

Let us define the corresponding cutted diagrams by
${\mathcal A_1}^{(1,1)}={\mathcal A_1}^{(2,2)}= {\rm i}\lambda$, , ${ \mathcal A_2}^{(1,2)}
={\mathcal A_3}^{(2,2)}=2 {\rm i}\lambda$, and 
${ \mathcal A_2}^{(2,2)}=3 {\rm i}\lambda$. 

Now, the first and third processes give the correct cutting equation, being
\begin{eqnarray}
&&2{\rm Im}\mathcal{M}_1=\frac{ \lambda^2}{2} \int\frac{d^4 q_1 }
{(2\pi)^4} \frac{d^4 q_2 }{(2\pi)^3}   \delta^4(p- q_1- q_2 ) \\ &\times& 
\left(  |{\mathcal A_1}^{(1,1)}|^2+|{\mathcal A_1}^{(2,2)}|^2     \right) \nonumber \;,
\end{eqnarray}
and 
\begin{eqnarray}
2{\rm Im}\mathcal{M}_3=\xi^2  \int\frac{d^4 q_1 }
{(2\pi)^4} \frac{d^4 q_2 }{(2\pi)^3}   \delta^4(p- q_1- q_2 )    |{\mathcal A_3}^{(2,2)}|^2    \;.\nonumber \\
\end{eqnarray}
The second process, due to the wrong sign of 
 $ {\rm Im} \mathcal M_2^{(1, 2)}(p)$, is responsible for a violation of unitarity, which however is very suppressed 
 in the delta coupling parameter of the order 
 of $\lambda^2\xi^4$. Hence, as long as the theory is understood to be effective one should have
 departures from unitarity only becoming relevant beyond the region of validity at which our 
 scalar theory has been defined.
 
For completeness we can explore alternative approaches to study perturbative unitarity.
Due to the presence of an indefinite metric and the ghost character of the field $\phi_2$, it is natural to explore whether the prescription that excludes negative metric states from the asymptotic Hilbert space allows restoring unitarity at one-loop level. 
By applying the Lee-Wick prescription, the only nontrivial
process is the first process in \eqref{amplitudes}.
However, since the cut diagram 
associated to the amplitud $ \mathcal M_1^{(2, 2)}(p)$ vanishes, in this case
there is also violation of unitarity 
of the order $\lambda^2$.
\section{Conclusions and outlook} \label{CONCL}
In this work, we have verified the delta theory's property of suppressing its radiative corrections beyond one-loop order. In particular, we have constructed the delta-theory associated to a scalar model with a quartic self-interaction term. We have quantized the model and find the corresponding propagators for the positive and negative metric fields.
We have tested unitarity at one-loop order by employing the optical theorem and the cutting equations. 
We have found suppressed violations of unitarity of the order of $\lambda^2\xi^4$
which provides a safe region to set up a meaningful effective theory based on 
the delta approach. The application of the Lee-Wick prescription, unfortunately increases
the order of violations of unitarity to the order $\lambda^2$. Regarding the effective point of view
and the richer structure provided by gauge models, we believe that 
relevant studies on perturbative unitarity for delta theories 
may come from analyzing the gravity sector, which we leave for future work.
\section*{Acknowledgments}
The authors are very grateful to Jorge Alfaro and Luis F. Urrutia for many discussions and comments.
R.A. has been supported by project Ayudant\'{i}a de Investigaci\'{o}n No. 352/1959/2017 
and 352/12361/2018 of Universidad del B\'{i}o-B\'{i}o.
C.M.R. acknowledges support 
by the research project Fondecyt Regular No.\ 1191553-Chile.



\begin{thebibliography}{99}
\bibitem{Delta_Gauge} 
 J.~Alfaro and P.~Labrana,
  Phys.\ Rev.\ D {\bf 65}, 045002 (2002).
\bibitem{BV}
I.~A.~Batalin and G.~A.~Vilkovisky,
Phys. Lett. B \textbf{102} (1981), 27-31; I.~A.~Batalin and G.~A.~Vilkovisky,
Phys. Rev. D \textbf{28} (1983), 2567-2582.
\bibitem{D-Alfaro}
J.~Alfaro,
``BV gauge theories,''
[arXiv:hep-th/9702060 [hep-th]].
\bibitem{DG}
  J.~Alfaro,
Phys. Part. Nucl. \textbf{44} (2013), 175-189; J.~Alfaro,
Phys. Lett. B \textbf{709} (2012), 101-105.
 \bibitem{DQ}
J.~Alfaro, P.~Gonzalez and R.~Avila,
  Class.\ Quant.\ Grav.\  {\bf 28}, 215020 (2011).
\bibitem{DE}
J.~Alfaro, M.~San Martin and J.~Sureda,
Universe \textbf{5} (2019) no.2, 51; J.~Alfaro and P.~Gonzalez,
AIP Conf. Proc. \textbf{1647} (2015) no.1, 80-88.
\bibitem{DM}
J.~Alfaro and P.~Gonzalez,
Class. Quant. Grav. \textbf{30} (2013), 085002
\bibitem{CF}
J.~Alfaro, C.~Rubio and M.~San Martin,
  [arXiv:2001.08354 [astro-ph.CO]].


\bibitem{Dirac}
P. A. M. Dirac,
``Bakerian Lecture -- The physical interpretation of quantum mechanics,''
Proc. Roy. Soc. A: Math., Phys. Eng. Sci. {\bf 180}, 1 (1942).




\bibitem{PU}
A.~Pais and G.~E.~Uhlenbeck, Phys.\ Rev.\  {\bf 79}, 145 (1950).


\bibitem{LW}
T.~D.~Lee and G.~C.~Wick,
 Nucl.\ Phys.\ B {\bf 9}, 209 (1969);
 T.~D.~Lee, G.~C.~Wick,
 Phys.\ Rev.\  {\bf D2}, 1033-1048 (1970).
 

\bibitem{Cut} 
 R.~E.~Cutkosky,
 J.\ Math.\ Phys.\  {\bf 1}, 429 (1960);  
 R.~E.~Cutkosky, P.~V.~Landshoff, D.~I.~Olive and J.~C.~Polkinghorne,
  Nucl.\ Phys.\ B {\bf 12}, 281 (1969).

\bibitem{G} 
B.~Grinstein, D.~O'Connell and M.~B.~Wise,
Phys.\ Rev.\ D {\bf 77}, 025012 (2008); J.~R.~Espinosa, 
B.~Grinstein, D.~O'Connell and M.~B.~Wise,
Phys.\ Rev.\ D {\bf 77}, 085002 (2008); J.~R.~Espinosa and B.~Grinstein,
Phys.\ Rev.\ D {\bf 83}, 075019 (2011).

\bibitem{Ap-LW}
L.~Modesto,
Nucl. Phys. B \textbf{909} (2016), 584-606.
 
  \bibitem{tree}  
  C.~M.~Reyes,
Phys.\ Rev.\ D {\bf 87}, no. 12, 125028 (2013); M.~Schreck,
  Phys.\ Rev.\ D {\bf 89}, no. 10, 105019 (2014); M.~Schreck,
  Phys.\ Rev.\ D {\bf 90}, no. 8, 085025 (2014).
  
  \bibitem{unitarity}
M.~Maniatis and C.~M.~Reyes,
Phys.\ Rev.\ D {\bf 89}, no. 5, 056009 (2014); J.~Lopez-Sarrion and C.~M.~Reyes,
  Eur.\ Phys.\ J.\ C {\bf 73}, no. 4, 2391 (2013).




  \bibitem{P-A}
D.~Anselmi and M.~Piva,
JHEP \textbf{06} (2017), 066; D.~Anselmi and M.~Piva,
Phys. Rev. D \textbf{96} (2017) no.4, 045009.


%
%
%





 

%
%
%
%
%


%
%
%
%
%
%
%
  



\end{thebibliography}
\end{document}